\documentclass[aps, superscriptaddress, twocolumn, floatfix, showpacs]{revtex4-1}

\usepackage{graphicx}
\usepackage{amsmath, amssymb}
\usepackage[usenames]{color}
\usepackage{soul}

\begin{document}

\newcommand{\comA}[1]{\textcolor{blue}{#1}}
\newcommand{\comC}[1]{\textcolor{red}{#1}}
\newcommand{\comR}[1]{\textcolor{ForestGreen}{#1}}

\newcommand\s{\dot{\gamma}}
\newcommand\sm{\dot{\gamma}_{mean}}
\newcommand\slo{\dot{\gamma}_{local}}
\newcommand\sa{|\dot{\gamma}|}

\newcommand\al{\textit{et~al.}\ }
\newcommand\ec{\textit{E. coli}}

\title{Living on the edge: \\  transfer and traffic of {\it E. coli} in a confined flow}

\author{N. Figueroa-Morales$^{1,2}$, G. Mi\~{n}o$^3$, A. Rivera$^2$, R. Caballero$^2$ \\
E. Cl\'{e}ment$^1$, E. Altshuler$^2$ and A. Lindner}

\affiliation{PMMH, UMR 7636 CNRS-ESPCI-Universit\'es Pierre et Marie Curie and Denis Diderot, 10, rue Vauquelin, 75231 Paris Cedex 5, France\\
$^2$ ``Henri Poincar\'e'' Group of Complex Systems and Superconductivity Laboratory, Physics Faculty-IMRE, University of Havana, 10400 Havana, Cuba\\
$^3$ Environmental Microfluidics Group, MIT, U.S.A. }

\date{\today}

\begin{abstract}
We quantitatively study the
transport of {\it E. coli} near the walls of confined microfluidic channels,
and in more detail along the edges formed by the interception of two
perpendicular walls. Our experiments establish the connection between
bacteria motion at the flat surface and at the edges and demonstrate the
robustness of the upstream motion at the edges. Upstream migration of {\it E.
coli} at the edges is possible at much larger flow rates compared to motion at the flat surfaces. Interestingly, the bacteria speed at the edges mainly results from collisions between bacteria moving along this single line. We show that upstream motion not only takes place at the edge but also in an ``edge boundary layer'' whose size varies with the applied flow rate. We quantify the bacteria fluxes along the bottom walls and the edges and show that they result from both the transport velocity of bacteria and the decrease of surface concentration with increasing flow rate due to erosion processes. We rationalize our findings as a function of the local variations of the shear rate in the rectangular channels and hydrodynamic attractive forces between bacteria and walls.

\end{abstract}

\maketitle

\section{Introduction}

The question of motility and transfer of microorganisms in their
environment is at the center of numerous issues. In many practical situations as in porous or fractured media, the pore size or the gap left for the micro-organisms to move can be very small and in theses confined situations surfaces become predominant. In addition, internal walls of porous media, as well as biological conducts like blood vessels, lymphatic ducts, urinary and reproductive tracks, are not simply perfect cylinders, but their surfaces have irregularities, as grooves and crevices, that make them different from simple regular surfaces.
The understanding of bacteria motion along complex surfaces
in the presence of flows is thus of strong importance for the control of microorganism transport in underground water resources, catheters or biological conducts.

\begin{figure}[!htb]
     \begin{center}
    \centering      \includegraphics[angle=0, scale=0.5]{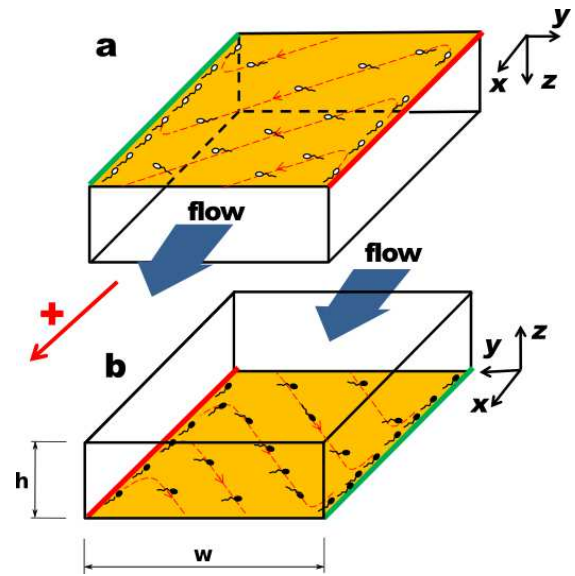}
     \end{center}
 \caption{Sketch of the experimental configuration. A rectangular microfluidic channel of height $h=20 \mu m$ and width $w=200 \mu m$ is observed using an inverted microscope. A bacteria suspension is flown at a volumetric rate $Q$ in the direction indicated by the blue arrows. The red arrow in the $x$ direction is our positive reference. Panels (a) and (b) indicate schematically \emph{typical} bacteria motion at the top and bottom surfaces, respectively, as well as at their interceptions with the lateral edges. Edges with a green (red) color picture correspond to mostly ``in-going'' (``out-going'') mean transverse flux.}
 \label{fig:sketch}
\end{figure}

Bacteria transport has been investigated in bulk flows and Marcos \al \cite{Marcos2012} have recently shown that bacteria drift with respect to the shear plane in Poiseuille flows. The drift has been explained as a consequence of the interaction between the chiral flagella of the bacteria and the shear rate \cite{Marcos2012, Fu2009}. As a consequence bacteria drift in opposite directions in the upper and lower half of a Poiseuille flow, induced by the opposite signs of the shear rate components.

The presence of walls modifies bacteria motion even without flow. In confined environments, bacteria are known to be attracted by flat surfaces and several studies have shown that the concentration of bacteria is significantly larger at the top and
bottom walls of square microchannels compared to the concentration
in the bulk \cite{Berke2008, Frymier1995, Drescher}.

The attraction towards the walls results from hydrodynamic interactions between swimming bacteria and walls. Typically an autonomously swimming bacterium can be seen as a force dipole. Then, bacteria described as "pushers'' \cite{Drescher} are attracted to their specular hydrodynamic image close to a solid wall \cite{leal2007advanced, Berke2008, DiLeonardo2011,chilukuri2014impact}. Furthermore, near a surface the rate of tumble has been observed to decrease with respect to the bulk \cite{molaei2014failed} also contributing to the long time bacteria spend very close to solid surfaces \cite{Drescher, schaar2014detention}. Bacteria motion at the surface is also modified compared to motion in the bulk due to lubrication forces between bacteria and walls. The viscous drag felt by the bacteria slows down the bacteria velocity \cite{DiLeonardo2011} and leads to the existence of circular trajectories due to their body rotation \cite{Frymier1995,Berg2004,Lauga2006}. Note that a purely kinetic approach \cite{Ezhilan2015, Li2009accumulation}, not taking hydrodynamic interactions with walls into account, also predicts increased bacteria concentrations at walls in confined geometries.

Under flow, the interaction between local shear and swimmer motion close to the surface results in upstream migration at small shear rates and downstream swimming at a given angle with respect to the flow direction at higher shear rates, reported for \textit{E. coli}  in \cite{Kaya2012} and for mammalian sperm in \cite{kantsler2014}. More generally, recent studies suggest that upstream swimming  takes place above a given threshold in shear rate for any front-back asymmetric micro-swimmer interacting hydrodynamically with a surface \cite{tung2015emergence}.

Concentration profiles stay flat in the bulk with a strong increase of concentration at the surfaces for small applied shear rates \cite{Gachelin2013}, but for higher shear rates more complex concentration profiles are observed and predicted in the direction of
the channel height \cite{Rusconi2014, Ezhilan2015, chilukuri2014impact}. Ezhilan and Saintillan \cite{Ezhilan2015}  and Chilukuri \al \cite{chilukuri2014impact} predict a decrease of the surface concentration
with increasing shear rate, but no experimental investigation of this phenomenon has been performed so far.

Less work has been devoted to the study of bacteria motion at wall interceptions. Bacteria concentration have been found to be even higher at these edges compared to flat surfaces \cite{Altshuler2013} and under flow bacteria motion is observed at the edges in a predominantly upstream way \cite{Hill2007,Altshuler2013} over long distances. This upstream motion is at the origin of anomalous
reconcentrations observed to be closely linked to the specific details of the confining
structure \cite{Altshuler2013,Figueroa2013}.

Here we study the behavior of bacteria while swimming in the vicinities and along the edges
resulting from the interception between bottom and lateral walls of confined rectangular
microchannels, as a first approach to understand their behavior in
response to shear in confined irregular structures.

We systematically quantify the
concentrations and velocities of bacteria at the horizontal surfaces
 and at the edges, as a function of the applied mean shear rate. We show that concentrations decrease exponentially with shear rate due to erosion of bacteria. The slower decay of bacteria concentrations at the edges can be explained by the smaller local shear rate experienced by bacteria at the edges compared to the surfaces, as well as strong attractive hydrodynamic interactions and suppression of bacteria tumbles. We establish the link between bacteria motion at the horizontal surfaces and the anisotropic bacteria concentrations observed at opposite edges. We show that the bacteria navigation along the edges takes place at speeds mainly given by collisions between bacteria moving along a single line along the edges. In contrast, at the bottom and top surfaces bacteria are transported downstream at a speed proportional to the mean shear rate, as soon as a critical shear rate is overcome.

We define and measure an order parameter that allows to
quantify the strong tendency of {\it E. coli} to navigate upstream
along the edges. It shows a transition from a symmetric mix of
downstream/upstream navigation at very low mean shear rates, to pure
upstream navigation at larger mean shear rates. We identify an ``edge boundary layer'' close to the edges, where bacteria also navigate upstream.

Finally, we show that the overall bacteria transport results from the dependency of both the concentrations and transport velocities on the mean shear rate.

\begin{figure*}
\begin{minipage}[b]{\textwidth}
 \begin{tabular}{l l l l}
   \textbf{a} &  &  & \textbf{c} \\
   \includegraphics[scale=0.5]{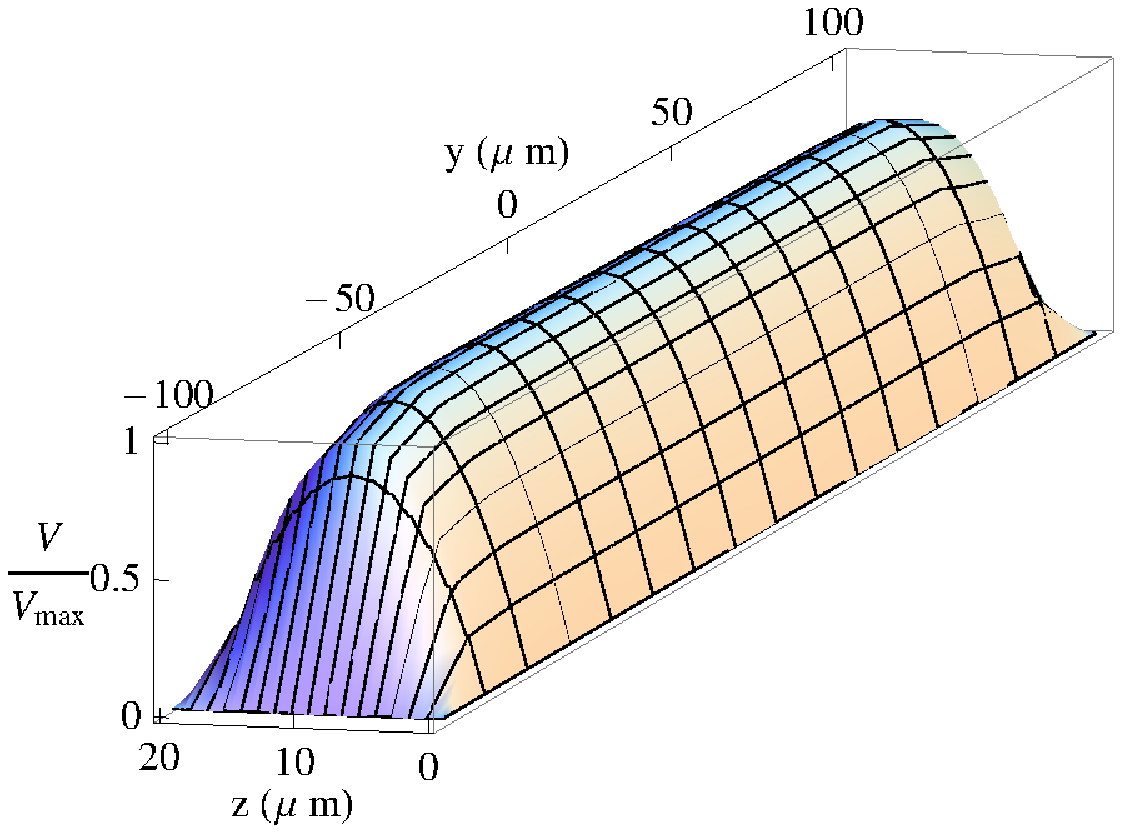} &
   \includegraphics[scale=0.5]{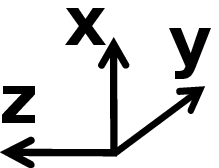} &  &
     \includegraphics[scale=0.5]{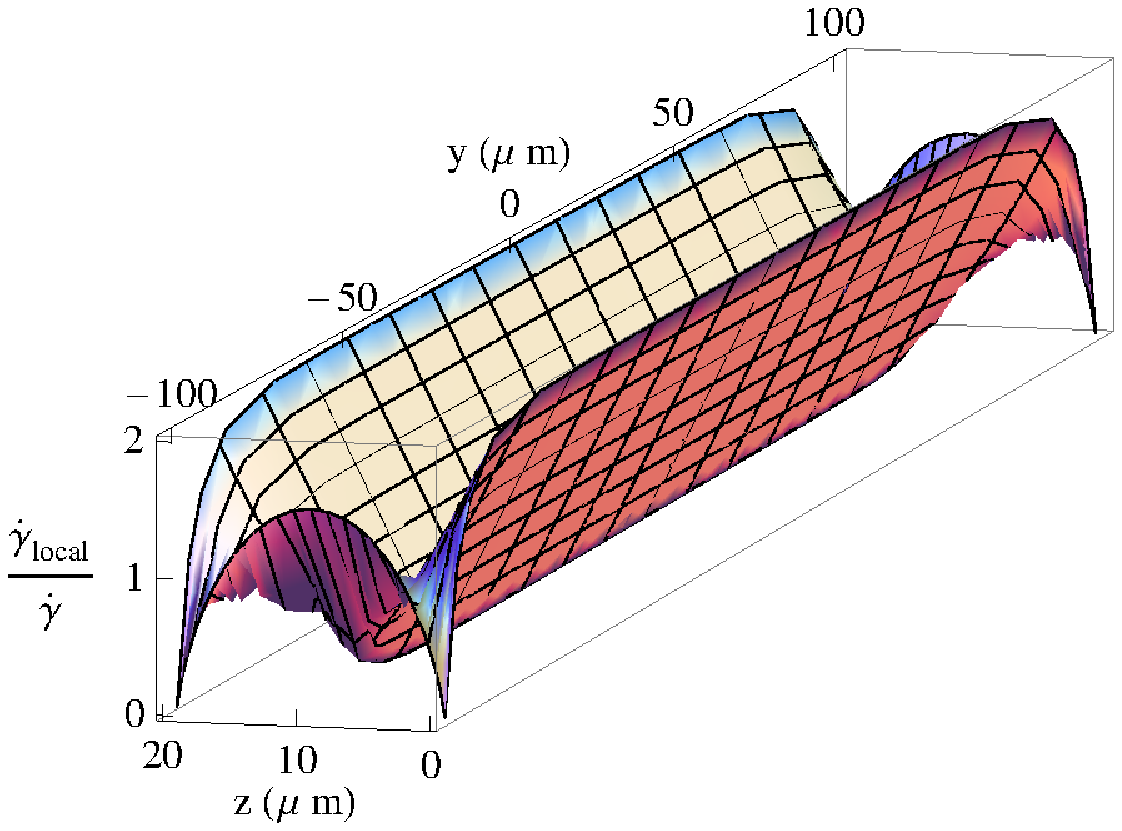} \\
   \medskip
   \textbf{b} &  & \textbf{d} \\
   \includegraphics[width=5cm,keepaspectratio]{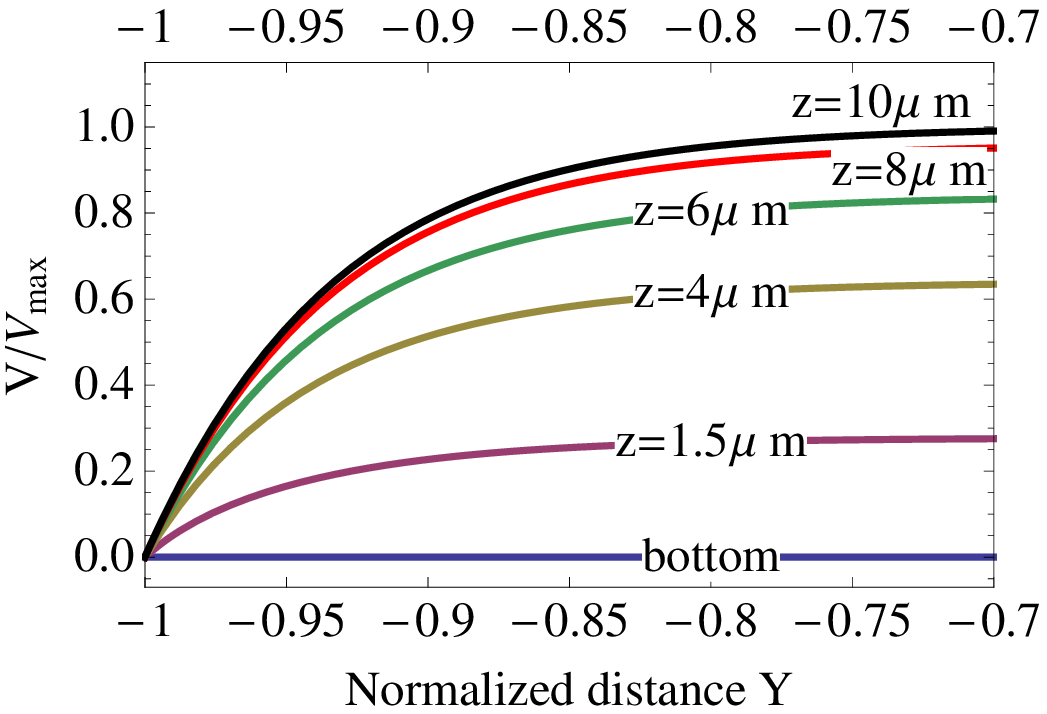} &  &  &
   \includegraphics[width=5cm,keepaspectratio]{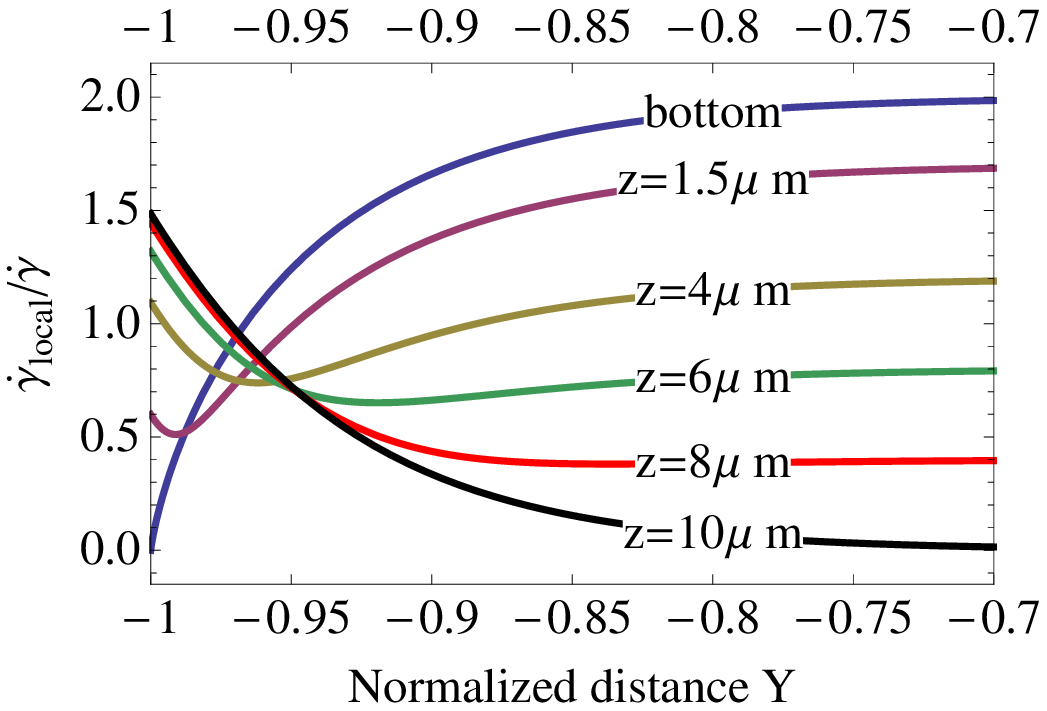}
 \end{tabular}
\end{minipage}
\caption{Flow velocities and shear rates for a rectangular channel ($w/h=10$) from the solution of Stoke's equations (eqn. \ref{sol_veloc}). (a) Normalized velocity field $V(y,z)/V_{max}$. (b) Zoom of normalized velocity profiles close to the lateral edges at different z positions versus distance normalized by the channel half-width $w/2$. (c) Normalized shear rate amplitude $\s_{local} (y,z)/\s$. (d) Zoom of the normalized shear rate profiles close to the lateral edges at different z positions versus distance normalized by the channel width $w$.}
  \label{fig:V_and_shear_profiles}
\end{figure*}

\section{Experimental set-up and methods}
\label{sec:experimental_details}

\textsl{Bacteria suspension~} -- The bacteria are wild-type {\it E. coli} (ATCC 11105).
Suspensions are prepared using the following protocol:  10 $\mu L$ of bacteria were grown overnight in 15 mL of a rich culture medium (LB). From this, 5 mL was washed using 25 mL of PBS and centrifuged. Thereafter, the pellet was re-suspended into Minimal Motility Medium (MMA)\cite{Minamino2003} and supplemented with K-acetate (0.34 mM) and polyvinyl pyrolidone (PVP: 0.005$\%$). After incubating for an hour in this medium to obtain a maximal activity, it was mixed with Percoll (1:1) to avoid bacteria sedimentation.  This controlled environment promotes motility but does not allow bacteria replication. The experiments where performed at $(25\pm 2)$ $^\circ C$. Under these conditions, the average swimming speed (far from the surfaces) is $28.7\pm~1.3 \mu m/s$. \textit{E.coli} perform run an tumble motions which can be regarded as a directed random walk process \cite{Berg2004, TorqueBerg}. At short times, the trajectory is ballistic with a swimming velocity $v_0$ and at long times, the motion is diffusive with a translational diffusion coefficient: $D = \frac{1}{6}v_0^{2} \tau$, where $ \tau $ is the cross-over time scale expressing the loss of directional orientation and is $\tau = 0.5 s$ for our experiments \cite{Gachelin2013, Matias2015}.

For the present experiments, the average bulk bacteria concentration is kept at $n_b = (3 \pm 0.5) \times 10^9$ bact./mL ($3 \times 10^{-3}$ bact.$/\mu m^3$) unless otherwise stated. The volume fraction $\phi$, based on a body volume of $1\mu m^3$ is $\phi=0.003$ and corresponds to a dilute regime.\\

\textsl{Microfluidic channel} -- The experimental cell is a rectangular channel made in PDMS using soft lithography techniques. The channel is $h = 20~\mu m$ deep, $w=200~\mu m$ wide and several millimeters long (see Fig. \ref{fig:sketch}). Confinement of bacteria suspensions by two parallel walls can be quantified by comparing the distance between the two walls (the channel depth) to a typical distance over which bacteria swim between two successive tumbles, yielding the confinement parameter $C = \frac{v_0 \tau}{h}$ \cite{Ezhilan2015}. For $C>1$ a bacterium crosses the channel depth essentially without changing direction and the walls play an important role in the bacteria dynamics. On the other hand, for $C<<1$ the motion of bacteria within the channel depth is diffusive and the channel walls only play a minor role. For our experimental system, we find $C=0.75$ and we are thus in a  situation where walls play a non-negligible role. \\

\textsl{Experimental protocol and bacteria detection} --  The suspension is observed using an inverted microscope (Zeiss-Observer, Z1) with a high magnification objective $100\times$ (field-depth 6 $\mu m$). Videos are taken with a digital camera Photron FastCam SA3 (1024 $\times$ 1024 $pixels$ or 158 $\times$ 158 $\mu m$) at a frame-rate of 500 fps unless otherwise stated. The bacteria suspensions were seeded with latex beads ($d=2~\mu$m, Beckman Coulter, density $\rho = 1.027 g/mL$ at a volume fraction $10^{-5} \%$) and injected inside the micro-channel at different flow rates $Q$ obtained by gravity over-pressure.
The focus position is set at a height immediately above the bottom surface where a first homogeneous layer of moving bacteria can be detected. At this position the cell bodies appear as white areas surrounded by a dark halo. Through image post-processing they are detected via their local intensity maximum. Then using a calibrated criterion for the maximal intensity, we quantified that bacteria detected in this way where within a distance  $z_0= 1.5 \mu m $ from the surface. Furthermore, we define that bacteria belong to {\it the horizontal surfaces} when they can be tracked in focus for at least 1 second, which corresponds to a traveling distance without flow of typically their own size including the flagella bundle.

Although the field depth of the microscope is quite small ($6 \mu m$), beads can be observed over the whole channel depth when flowing through the micro-channel. The fastest bead in each video is used to determine the maximum flow velocity $V_{max}$ in the channel. The uncertainty of this method depends on the number of observed beads and is almost negligible as soon as we observe more than 10 beads. It can however be significant at very low flow rates. Maximal velocities detected vary between 0 and $4150\mu m/s$.

\section{Velocity and shear profiles }
\label{section:rectangular channel}

The rectangular channel we use here has an aspect ratio $w/h=10$. Velocity profiles differ thus from an ideal parabolic Hele-Shaw flow profile, where the velocity is invariant in the transverse direction $y$, for distances to the lateral wall larger than the cell height $h$. This will influence the local values of  flow velocities and shear rates and thus the transport properties of bacteria at the surfaces. We present in Fig. \ref{fig:V_and_shear_profiles} the flow velocities and shear rates computed for a rectangular channel with the same aspect ratio as our experimental channel. Velocities and shear-rates are normalised respectively by the maximal velocity $V_{max}$ and by the average shear rate $\dot{\gamma} =\frac{2 V_{max}}{h}$.

The amplitude of the local shear rate  is defined as:
$\s_{local}(y,z)= \sqrt{ ( \frac{\partial V}{\partial y}(y,z))^2 + ( \frac{\partial V}{\partial z}(y,z))^2}$. General views and zooms of the longitudinal velocities and shear rates are displayed on  Figs. \ref{fig:V_and_shear_profiles}(a)(c) and Figs. \ref{fig:V_and_shear_profiles}(b)(d) respectively. The zooms show the regions close to the edges that we are specifically interested in. The flow and shear rate profiles are obtained from an exact solution of the Stokes equation for the longitudinal velocity field $V(y,z)$ with no-slip boundary conditions \cite{tabeling2010introduction}:
\begin{eqnarray}
 V(y,z) =  \sum_{n,odd}^\infty \frac{4 h^2 c}{\pi^3 n^3 }
 \left[
1- \frac{\cosh\left(\frac{n \pi y}{h}\right) }{\cosh\left( \frac{n \pi w}{2h}\right)}
\right]
\sin\left( \frac{n \pi z}{h}\right)
\label{sol_veloc}
\end{eqnarray}
where $c=-\frac{\nabla p}{\mu}$, (with $\mu$ the dynamic viscosity and $p$ the pressure), $-\frac{w}{2}<y<\frac{w}{2}$ and $0<z<h$.
Interestingly, one can see that close to the edges there is a significant variation of the hydrodynamic conditions compared to the situation at the bottom/top surfaces. At a typical distance from the bottom wall, corresponding to our observation plane ($z \approx 1.5 \mu m$), we find a velocity reduction ($V_{surface} \approx 0.3 V_{max}$) and an increase in the local shear rate ($\s_{surface} \approx 1.7 \s$), while at the same height and $1\mu m$ from the lateral wall, corresponding to the edge, we find $V_{edge} \approx 0.05 V_{max}$ and $\s_{edge} \approx 0.5 \s$.

 \section{Results and discussion}

\subsection{Experimental observations}
\label{subsection:observations}

In absence of flow, a large amount of bacteria swim close to the
bottom surface and we observe them performing circular
trajectories (not shown) in agreement with previous observations \cite{Frymier1995,Berg2004,Lauga2006}.

\begin{figure}[!htb]
     \begin{center}
          \includegraphics[width=0.85\linewidth]{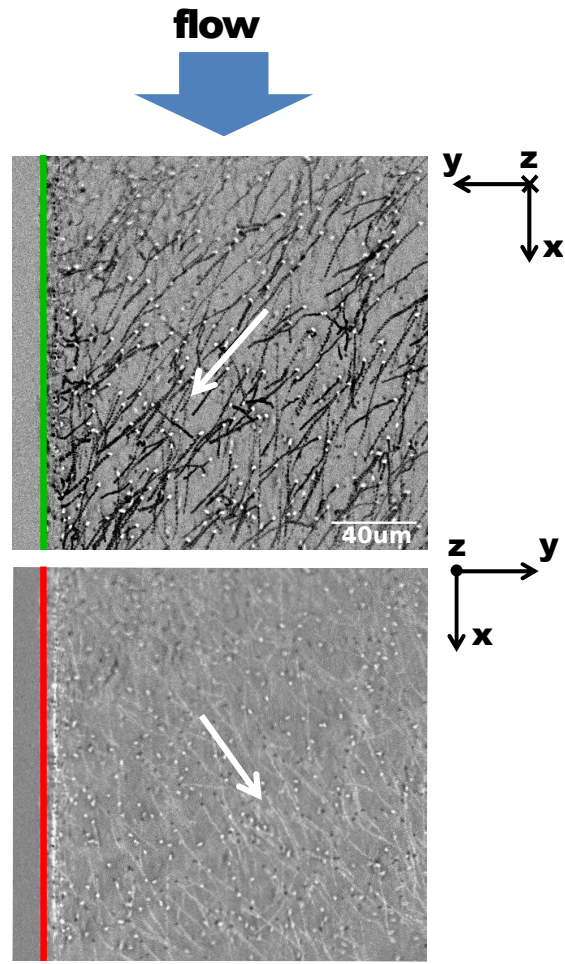}\\
     \end{center}	
 \caption{Surface rheotaxis. Superposition of snapshots for the top and bottom channel surfaces.
 Individual bacteria trajectories in the channel are visible from image superposition and white arrows indicate the average direction of bacteria motion. The green and red lines indicate respectively ``in-going'' and ``out-going'' edges, analogous to the ones sketched in Fig. \ref{fig:sketch}. For these snapshots the bulk bacteria concentration of the reservoir was $(4 \pm 0.5) \times 10^9$ bact./mL. The videos were taken at $40\times$ magnification and 30 fps with a digital camera PixeLINK PL-A741-E ($512 \times 512$ pixels or $197 \times 197 \mu m$), at a mean shear rate of $19 s^{-1}$.}.
 \label{fig:rheotaxis}
\end{figure}

When a flow is applied, the scenario changes radically. The swimming direction is modified and at very small shear rates we observe upstream motion of bacteria, as also reported by Kaya and Koser \cite{Kaya2012}. At higher shear rates (corresponding to most of our experiments) bacteria navigate diagonally downstream (left or right as a function of the sign of the shear gradient), see Fig. \ref{fig:sketch} and snapshots on Fig. \ref{fig:rheotaxis}.
This effect of ``surface rheotaxis'' is also in agreement with previous observations in \cite{Kaya2012}.

\begin{figure}[!htb]
     \centering
     \includegraphics[angle=0, scale=0.55]{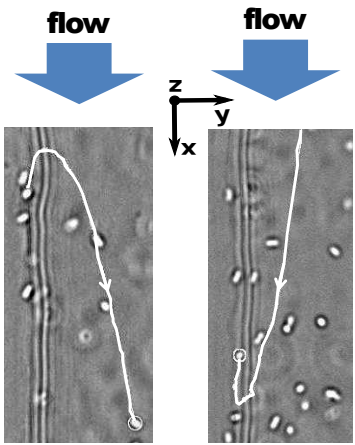}
 \caption{ Attachment/detachment processes visualized at the edge. The background ($25 \times 50 \mu m$) snapshots are taken near one lateral edge between the bottom surface and a lateral wall. The white line trajectories superimposed on the picture corresponds to bacteria detaching from the edge (left) and attaching to it (right). Mean shear rate is $\dot{\gamma}=220 s^{-1}$, (the edge corresponds to an ``in-going'' case, but it was not colored in green for clarity).}
 \label{fig:attachment_detachment}
\end{figure}

 As bacteria reach an interception between the top and bottom surfaces with the vertical walls, they reorient along the edges, leading predominantly to an upstream motion. Due to shear, bacteria can also be eroded from the edges. In Fig. \ref{fig:attachment_detachment}, we present two typical trajectories corresponding to attachment (left panel) and detachment (right panel) from an edge. Note that for these trajectories bacteria swim indeed upstream at the edge.

 It is important to notice that surface rheotaxis has a strong influence on the balance of fluxes concerning the bacteria traffic at the 4 edges, as it breaks their geometrical symmetry. This is why we make a distinction between edges corresponding to ``in-going'' (green line) and ``out-going'' (red line), represented in Figs. \ref{fig:sketch} and \ref{fig:rheotaxis} and that will be discussed further in section \ref{subsec:concentrations}.

 \subsection{Surface {\it vs.} edge erosion}
\label{subsec:concentrations}

\begin{figure}[!htb]
     \begin{center}
   \flushleft \textbf{a} \\
   \centering       \includegraphics[width=0.85\linewidth]{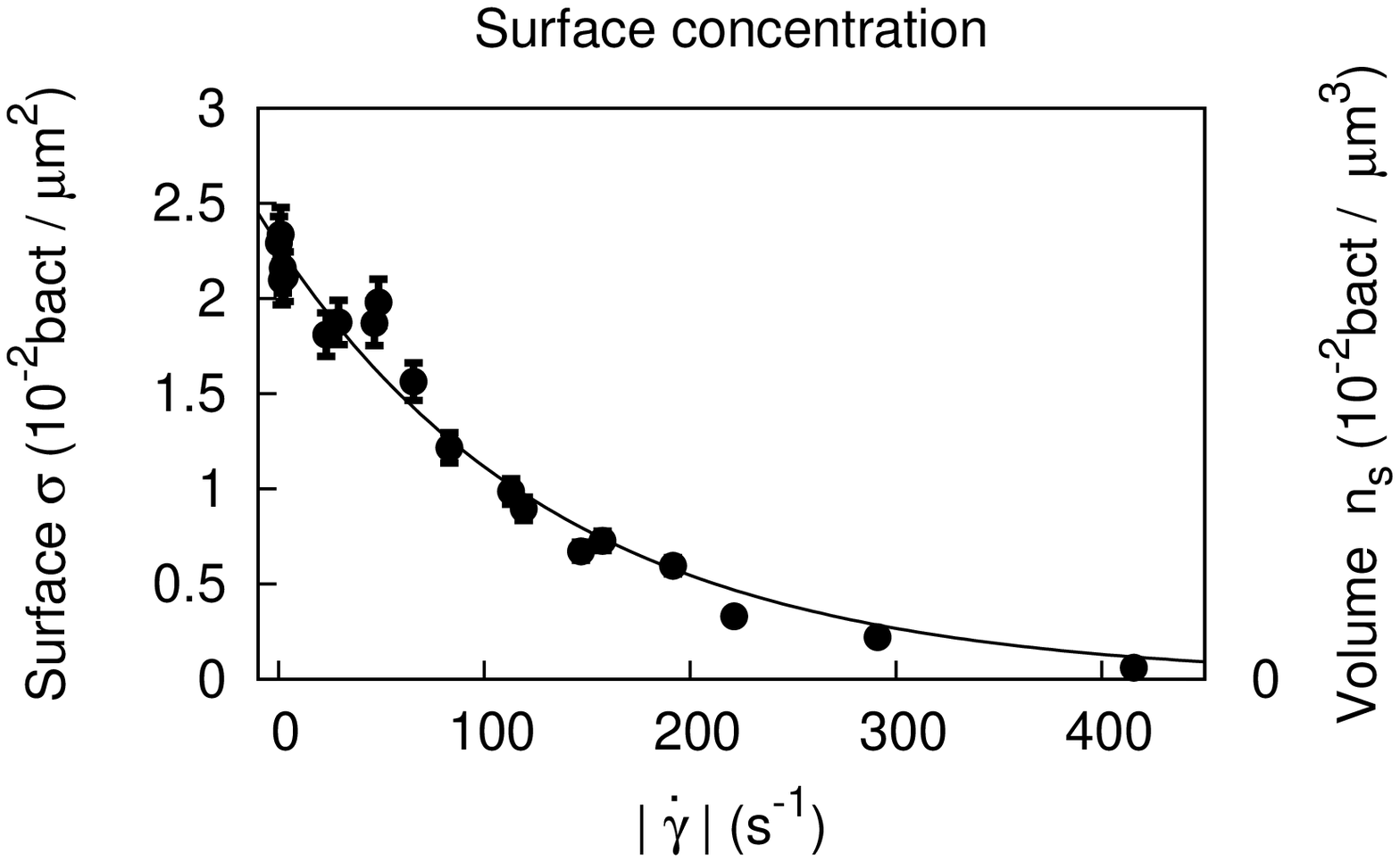}

     \flushleft \textbf{b}\\
 \centering    \includegraphics[width=0.85\linewidth]{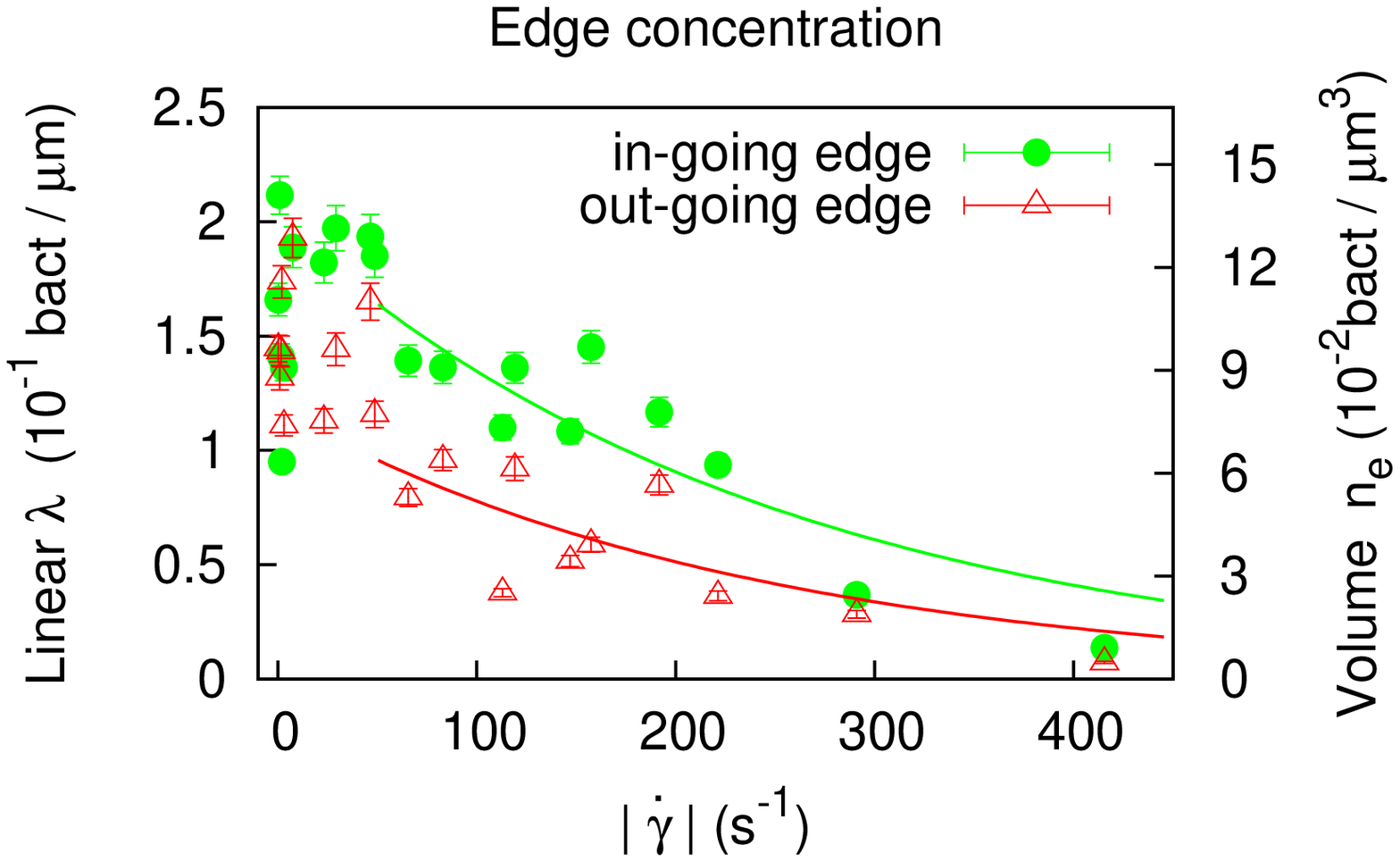}
		\end{center}
 \caption{Surface erosion: bacteria concentrations vs absolute shear rate value $|\dot{\gamma}|$ . Lines are exponential fits: $A \exp{(-|\dot{\gamma}|/\dot{\gamma}^0)}$. (a) Bacteria concentrations are measured at the channel bottom surface, excluding $10\mu m$-wide stripes from the lateral walls. $A_s=(2.81\pm0.12) 10^{-2} bact/\mu m^2$ and $\dot{\gamma}^{0}_s=(143\pm20)s^{-1}$.  (b) Linear concentrations at the edges for ``in-going'' and ``out-going'' edges . Fit parameters: $A_i=(2.0\pm0.7)10^{-1}bact/\mu m$ and $\dot{\gamma}^{0}_i=(250\pm 75)s^{-1}$ (in-going) and $A_o=(1.2\pm0.7)10^{-1} bact/\mu m$ and $\s^{0}_o=(240\pm 100)s^{-1}$ (out-going).}
 \label{fig:n}
\end{figure}

Here we quantify the evolution of the bacteria concentration as a
function of the mean shear rate at the flat horizontal surfaces and the edges. Fig. \ref{fig:n}
shows the mean bacteria surface concentration at the bottom wall and at the edges of
the channel as a function of the mean shear rate.

In Fig. \ref{fig:n} the surface concentration $\sigma$ is defined as the number of bacteria observed within an area S of the surface, and within a distance $z_0=1.5 \mu m$ from it, divided by S. The volume concentration $n_s$ near the surface is then given by the number of bacteria observed in the same region, but divided by the volume $z_0S$. The linear edge concentration $\lambda$ is defined as the number of bacteria observed along a distance $L$ of the edge, divided by $L$. Finally, the volume concentration near the edge $n_e$ is given by the edge linear concentration divided by $1.5\mu m^2$, i.e., we are assuming that bacteria swim along a corridor of width $1\mu m$ and height $1.5\mu m$ parallel to the edge. At zero shear rate, the volume concentrations for the surface and the edges shown in Fig. \ref{fig:n}(a) and (b) are found to be around $1.5 \times 10^{-2}$ and $10^{-1}$ bact$/\mu m^3$ respectively. Compared to the ``bulk'' volume concentration of the reservoir of ($n_b = 3\times 10^{-4}$bact$/\mu m^3$) the concentration is thus 5 times higher at the flat surfaces and 30 times higher at the edges. Ezhilan and Saintillan \cite{Ezhilan2015} predict a typical increase of the surface concentration at a flat surface for parameters comparable to our experimental conditions of typically 2-3 times. A purely kinetic model as theirs might thus not be enough to explain the large increase in concentration observed in our experiments. This indicates the importance of hydrodynamic attractive forces \cite{chilukuri2014impact} (or others) to maintain bacteria near the walls for long times.

We now discuss the evolution of concentrations when a flow is applied. From Fig. \ref{fig:n}(a) one can see that with increasing shear rate, the surface concentration decreases strongly, indicating erosion of bacteria from the wall. The concentration decrease is well described by an exponential decay as $A \exp{(-\sa/\s_0)}$ with $\s_0=(143\pm20)s^{-1}$. This typical erosion shear rate is large compared to the inverse of a typical hydrodynamic time scale for swimming bacteria, that can be defined as $v_0/l$, where $v_0$ is the bacteria swimming speed and $l$ is a typical bacterium size. This difference might be explained once again by hydrodynamic interactions between the bacteria and the surface.

The viscous drag on a bacterium can be estimated as $F_e \propto \mu l^2 \s $, with $\mu$ the fluid viscosity and where modifications of the viscous drag due to the presence of the wall are neglected. On the other hand, the hydrodynamic attraction force in the presence of a flat surface scales as $F_a \propto \mu l (\frac{l}{r_w})^2 v_0 $, with $r_w$ the distance between the swimmer and the wall and $l$ the dipole size, corresponding roughly to the bacteria body size (possibly including the flagella) \cite{leal2007advanced}.  From the balance between $F_e $ and $F_a$, a typical erosion shear rate can be obtained:
\begin{equation}
\s_0 \propto \left(\frac{l}{z_0}\right)^2 v_0/l.
\end{equation}
Therefore, a distance of $z_0 = 1 \mu m$, $l = 10 \mu m$ and $v_0= 30 \mu m/s$ yields: $\s_0  \approx 3\times10^2 s^{-1}$,  which is about the right order of magnitude for erosion, as seen in Fig. \ref{fig:n}(a). However, this estimation is critically sensitive to the choice of the ratio $\frac{l}{z_0}$ and clearly, should be validated on a more refined hydrodynamic model, possibly also including the kinetic contribution of bacteria incoming from the bulk flow.

Fig. \ref{fig:n}(b) shows the linear bacteria density at the two edges corresponding to the bottom
of the channel as a function of the mean shear rate.  An important observation can be made from this figure. The concentrations at the right and left edges are identical only at very small or very high shear rates. With increasing shear rate the concentration at the in-going edge increases whereas the concentration at the out-going edge decreases, leading to an asymmetry between the two edges. This has already been qualitatively observed in \cite{Hill2007} and results from the bacteria transport at the flat surfaces. From the snapshots on Fig. \ref{fig:rheotaxis} it is clear that bacteria swim, in average, with a finite angle compared to the flow
direction. This brings bacteria preferentially towards a given edge (the in-going edge).
At the bottom surface bacteria are observed to drift towards the left with respect to the flow direction, leading to a concentration increase at the left edge, from the point of view of an observer moving with the flow. At higher shear rates a decrease of the bacteria concentration with shear rate is again observed. At very large shear rates the erosion (detachment) of bacteria is so strong that the concentration tends towards zero for both edges. We attempt to adjust this decrease also by an exponential, and even if the quality of the fit for the edges is less good compared to the flat surface it leads to an estimate of the decay rate, which is found, by separately fitting the two data sets, to be approximately $\s^{0}=(250\pm 100)s^{-1}$ for both edges.

So, the shear rate associated with the concentration decrease is larger for the edges than for the surfaces and bacteria are thus eroded more slowly from the edges compared to the surfaces. In section \ref{section:rectangular channel} we have shown that the local shear rate at the edges is $3-4$ times smaller compared to the local shear rate at the flat surfaces. This difference in local shear rate is enough to account for the difference in $\s_0$ observed. We can however, within our experimental resolution, not exclude other effects that might make bacteria more resistant to erosion at the edges, as increased hydrodynamic attraction at the corner compared to the surface. Actually, the attractive interaction with the walls at the edges can be seen as stemming from the interaction between the bacterium and two specular images, each of them situated at the opposite side of each wall, plus a third specular image in a direction extrapolated from the segment going from the actual bacteria to the interception between the walls. Furthermore a smaller bacteria concentration at the flat surface for larger shear rates could also result in a smaller concentration at the edges, as less bacteria reach the edges reducing bacteria concentrations there.

 \subsection{Transport of bacteria by the flow}

\subsubsection{At the flat surfaces}

We start by reporting the bacteria velocities at the flat surfaces. First, without flow, we noticed a significant decrease of bacteria velocities close to the surfaces ($13\pm2 \mu m/s$) when compared with the velocities in the bulk ($28.7\pm1.3\mu m/s$).
Second, we report the transversal and longitudinal velocities of bacteria at the flat surfaces under flow.
Bacteria
move at the surfaces of the microchannel following transversal,
straight trajectories as is illustrated in Fig. \ref{fig:rheotaxis}.
By tracking individual bacteria, we were able to obtain their velocities and orientation. We project the velocities on the $x$ and $y$ axes to obtain the longitudinal and transversal bacteria velocities. The transversal velocity on the surface always points in the vorticity direction, indicating that bacteria move to the left with respect to the flow direction at the bottom surface. The longitudinal direction at the surface is referred to the fluid direction, being positive when bacteria move downstream and negative when they migrate upstream.

\begin{figure}[!htb]
 \flushleft \includegraphics[width=0.85\linewidth]{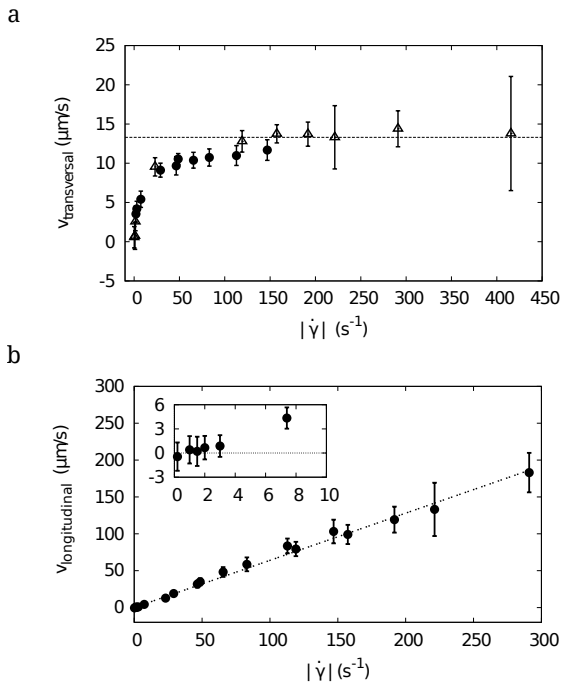}

 \caption{(a) Average transverse bacteria velocities at the horizontal surfaces as a function of the mean shear rate. Experimental data collected for both positive (filled circles) and negative (open triangles) flow directions at the same surface. The horizontal line indicates the average swimming velocity of bacteria at the surface at zero shear rate.  (b) Absolute value of the mean longitudinal bacteria velocities
for various flow rates $|\dot{\gamma}|$ at the bottom horizontal surface. The line is a linear fit to the data with a slope $\Lambda= 0.64 \pm 0.02 \mu m$.  In inset, zoom on the longitudinal velocity $<v_x>$.
  \label{fig:vy}}
\end{figure}

Fig. \ref{fig:vy}(a) shows the mean transversal bacteria velocity, oriented
from the out-going to the in-going edge.
If the flow is reversed, the transversal velocity also reverses, and the in-going and out-going edges
exchange positions.  The different symbols correspond to different flow directions. As the shear rate increases, we
see a steep increase in the transversal velocity, until it saturates at a value that coincides with
the average swimming velocity of the bacteria at surfaces without flow. This means that bacteria swim
almost perpendicular to the flow, as reported by Kaya and Koser \cite{Kaya2012}. This saturation occurs for a mean shear rate of the order of
40~$s^{-1}$.

The bacteria moving at the surfaces feel a Stokes drag from the local flow and can thus be transported downstream. In Fig. \ref{fig:vy}(b), the mean longitudinal velocity far from the lateral walls is displayed as a function of the mean shear rate $\dot{\gamma}$. The effect of the flow is an entrainment proportional to $\dot{\gamma}$. Due to the fact that bacteria swim mostly in a direction perpendicular to the flow direction at higher mean shear rates the swimming speed does not influence the bacteria transport. The longitudinal bacteria velocity is found to be smaller than the local flow velocity ($v_{longitudinal}=0.2 V_{surface}$), which can be explained by hydrodynamic interactions such as lubrication forces slowing down the mean transport velocity. At very small shear rates bacteria can swim upstream and average bacteria velocities in a direction opposite to the flow direction have been reported by Kaya and Koser \cite{Kaya2012}. Here we explore much higher flow rates, and the mean entrainment direction is essentially along the flow. Only for the smallest applied flow rate (see inset \ref{fig:vy}(b)) negative bacteria velocities has been detected.

 \subsubsection{Upstream {\it vs.} downstream traffic at the edges}
 \label{subsec:edge_navigation}

\begin{figure}[!htb]

         \flushleft \textbf{a} \\
  \centering        \includegraphics[width=0.85\linewidth]{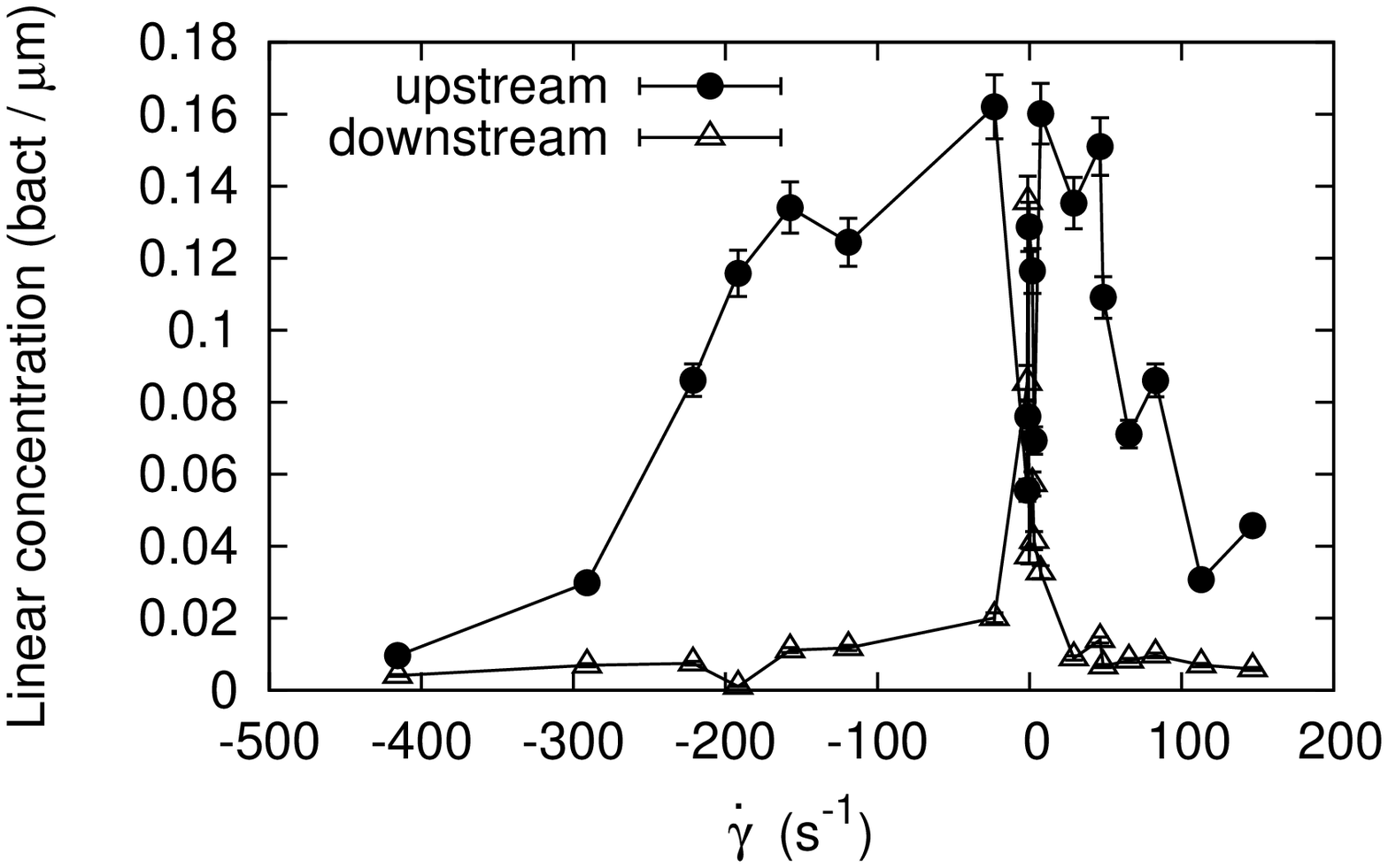}\\
          \flushleft \textbf{b} \\
  \centering        \includegraphics[width=0.85\linewidth]{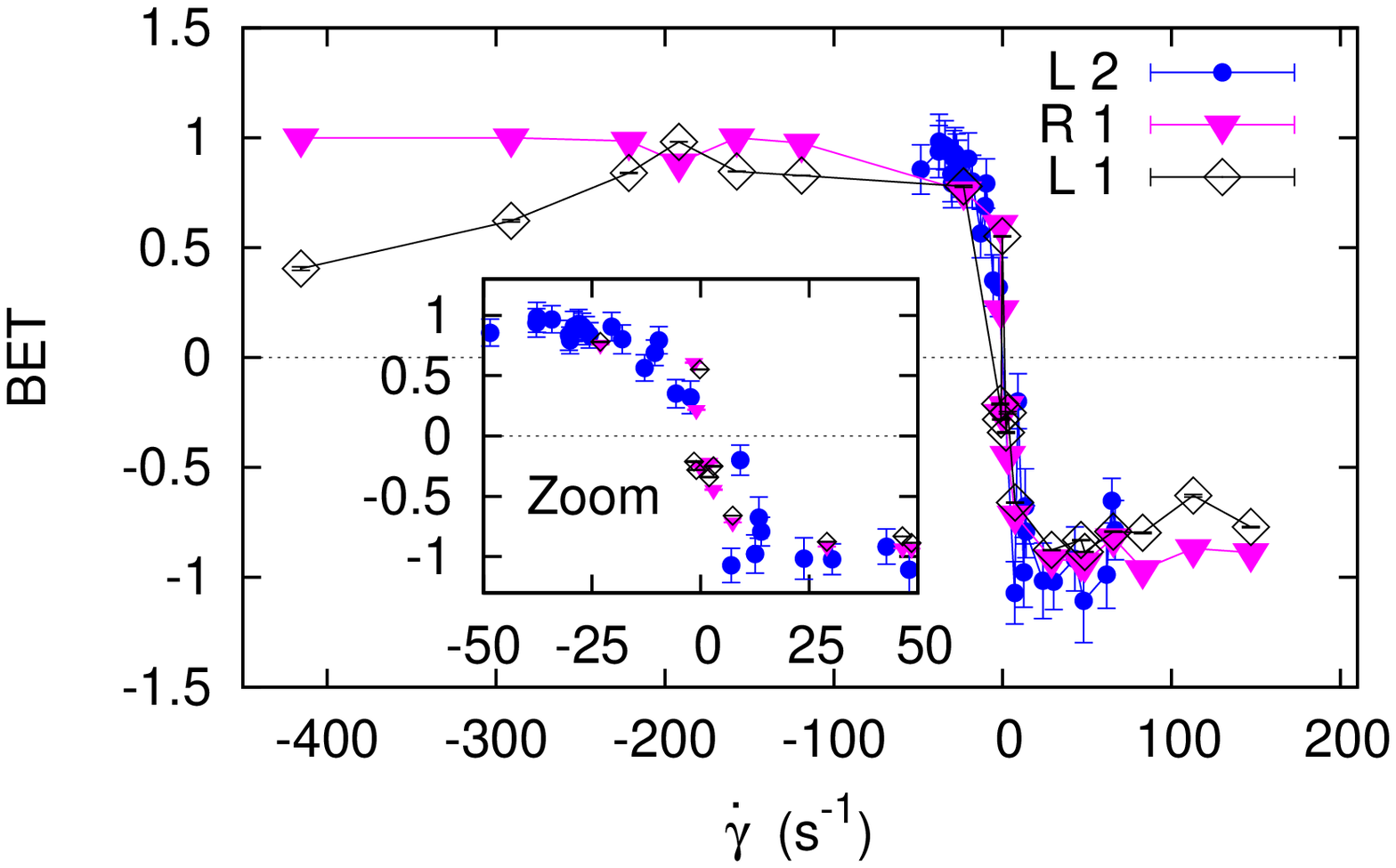}
          \flushleft \textbf{c}\\
   \centering       \includegraphics[width=0.85\linewidth]{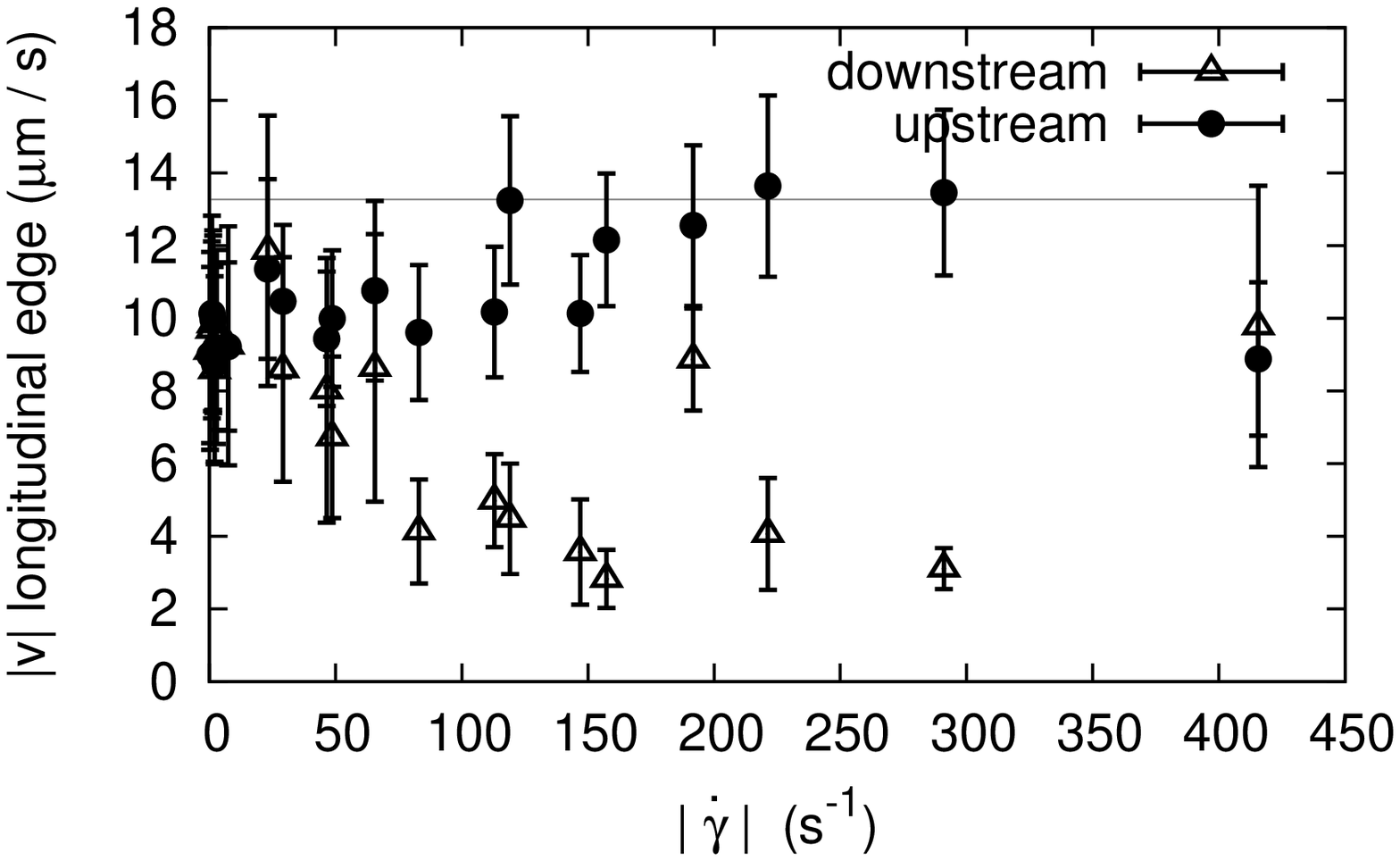}

 \caption{Bacteria transport at the edges. (a) Concentration of bacteria along the left edge. Note that the ``in-going'' edge for $\sm<0$ becomes an ``out-going'' edge when $\sm>0$.
(c) Bias parameter in Edge Traffic (BET) as a function of the shear rates (see Eq. \ref{BET}),
The meaning of the labels in the different symbols in the picture are: \textit{L1} and \textit{R1} Left and right walls respectively in an experiment at a bulk concentration in the reservoir of $(3 \pm 0.5) \times 10^8$ bact./mL; \textit{L2} left wall at $(4 \pm 0.5) \times 10^8$ bact./mL. (c) Absolute value of the mean longitudinal velocity for the bacteria population moving along the edges. The horizontal line is their mean velocity at the center of the channel at zero flow. }
 \label{fig:edge}
\end{figure}

In this section, we quantify in detail the traffic of {\it E. coli} moving along the edges.
Fig \ref{fig:edge}(a) displays the concentration of bacteria swimming
upstream and downstream at a given edge as a function of the mean
shear rate. Interestingly, the number of bacteria swimming upstream along the edges largely
dominates the number of bacteria swimming downstream. The slight asymmetry in the curves for positive and negative mean shear rates is due to the fact that the edge switches from an ingoing edge ($\s<0$) to an outgoing ($\s>0$) edge when the flow is inverted. The concentration of bacteria swimming downstream is observed to be very small as soon as the mean shear rate is larger than $|\s|\geq$ 25 s$^{-1}$. Shear induced orientation near the edges seems to be the cause of this difference: as bacteria approach the edge transversally as illustrated in Figs. \ref{fig:sketch} and \ref{fig:attachment_detachment}, the local shear rotates the bacteria body that aligns preferentially in the direction
facing the flow \cite{Hill2007,Altshuler2013}. This shear induced
orientation of bacteria does not take place at very small flow
rates, and the concentrations of upstream and downstream bacteria at
the edges are similar.

In order to further quantify the bias in the direction of the bacteria traffic along the edges, we define a  parameter called Bias in the Edge Traffic or BET number, as:
\begin{equation}
BET=\frac{\lambda_{up}-\lambda_{down}}{\lambda_{up}+\lambda_{down}}  \mbox{,}
\label{BET}
\end{equation}
where $\lambda_{up}$ and $\lambda_{down}$ are respectively the linear concentrations of
bacteria moving in the positive and negative directions relative to
the red arrow shown in Figs. \ref{fig:sketch} and
\ref{fig:rheotaxis}(a). BET thus allows to visualize the bias in the bacteria navigation without taking the total bacteria concentration or the erosion of bacteria from the edges into account.
Fig. \ref{fig:edge}(b) shows the dependence of BET on the mean shear rate $\dot{\gamma}$ for different experiments, i.e. different channels of similar geometry, different concentrations and measured at different edges.
When $\dot{\gamma}=0$, there is an equal amount of bacteria moving
up and downstream, so $BET=0$. As the shear rate increases in the
positive direction,
more and more bacteria move upstream (i.e., in
the negative direction), until $BET=-1$. A similar reasoning explains
the shape of the curve of Fig. \ref{fig:edge}(b) for negative shear
rates.

We now consider bacteria velocities at the edges. The average velocities of bacteria moving upstream and downstream at the two bottom edges are displayed on Fig. \ref{fig:edge}(c). Bacteria velocities are found to be identical for the two populations for very small shear rates. Velocities of about $10 \mu m/s$ (slightly smaller than the swimming speed at the flat surface) are observed. Surprisingly, with increasing shear rate upstream swimming bacteria become faster in average, whereas downstream swimming bacteria slow down. At even higher shear rates, the tendency is reversed and the downstream swimming bacteria see their velocity increased whereas the upstream swimming bacteria see their velocity decreased. These observations indicate transport dynamics very different from those reported at the flat surface, where the mean bacteria velocity is directly proportional to the shear rate (see. Fig. \ref{fig:vy}(b)). The transport velocities at the edges can thus not be simply explained by the hydrodynamic drag exerted by the local flow, but we associate these observations with the specific dynamics of bacteria moving along a unidimensional corridor along the edges.

In general, bacteria can move up and downstream at the edge, leading to collisions between swimmers. We have observed that during frontal collisions, bacteria slow down for a certain time until they cross each other and continue swimming along the edge. We have also observed bacteria to detach from the edge during such a collision. Similar slowing down or detachment can also take place during rear collisions between two bacteria swimming in the same direction, but are much less frequent.

For very small shear rates the concentration of bacteria at the edges is large, and collisions are frequent. These collisions are at the origin of the decreased transport speed of bacteria at the edges compared to the flat surfaces. As soon as the shear rate is larger than $25s^{-1}$ the number of bacteria swimming upstream is much larger than the number of downstream swimmers (Fig. \ref{fig:edge}(a)), resulting in more frequent collisions for bacteria swimming in the flow direction, and the downstream swimming population is thus more strongly slowed down, provoking a decrease of their velocity compared to the upstream swimming population.

With increasing flow rate, the concentration of bacteria at the edges decreases and bacteria transport velocities increase again, due to a decrease in the number of collisions.  At very high shear rates, only very little bacteria are left at the edges and we expect collisions not to play an important role any more. In this range of mean shear rates, the mean upstream bacteria velocities are observed to be close to the swimming speed measured at the flat surfaces without flow, indicating that bacteria swim upstream nearly undisturbed by the local flow. It is worth noticing that the speeds of individual upstream and downstream bacteria between collisions are always roughly similar and also very close to the bacteria swimming speed measured at the flat surfaces without flow.
For the downstream swimming bacteria the small average velocities measured in this range of shear rates are due to their attachment dynamics. During a typical attachment process under flow, a bacterium, advected by the main stream arrives at an edge, flips and starts swimming upstream. During this flipping process, the swimmer changes its velocity direction from downstream to upstream, contributing to the statistics of downstream swimmers velocity for a short lapse of time, with a velocity that decreases to zero.

At even higher shear rates the viscous drag on the bacteria might become important, leading to a decrease of the mean velocity of upstream swimming bacteria and a decrease for the downstream swimming bacteria. At a shear rate of $\s$=400$s^{-1}$ the local flow velocity at the edge is $20\mu m/s$. At the flat surfaces bacteria where transported downstream at velocities five times smaller compared to the local flow velocity. Assuming a similar relation at the edges this would correspond to a transport velocity of $4 \mu m/s$ that needs to be added to the swimming speed of the bacteria. This is not in contradiction with the observed decrease in average transport velocity of upstream swimming bacteria at these high mean shear rates. Most likely a correct modeling of the lubrication forces would lead to an even smaller expected transport velocity at the edges. The striking fact that bacteria move at the edges nearly unperturbed by the flow can thus at last partially be explained by the decreased local flow velocities at the edges compared to the flat surfaces. Increased drag close to a corner and the significant strength of the hydrodynamic interactions between bacteria and the edges might even enhance this effect.

As overall bacteria speeds at the edges vary little as compared to concentrations variations, the total bacteria flux at the edges is directly proportional to the concentration profile represented on Fig. \ref{fig:edge}(a) and has not been represented separately.

  \subsection{Edge boundary layer}
 \label{subsec:EBL}
The previous measurements show that the edges have singular transport properties, as there is a significant flux of bacteria moving against the flow. Here we characterize whether this upstream motion is restricted to the edges, by measuring the bacteria flux along the flow direction at the surfaces and close to the edges and identify what we call the Edge Boundary Layer (EBL).

\begin{figure}[!htb]
     \begin{center}
        \flushleft \textbf{a} \\
  \centering  	\includegraphics[width=0.85\linewidth]{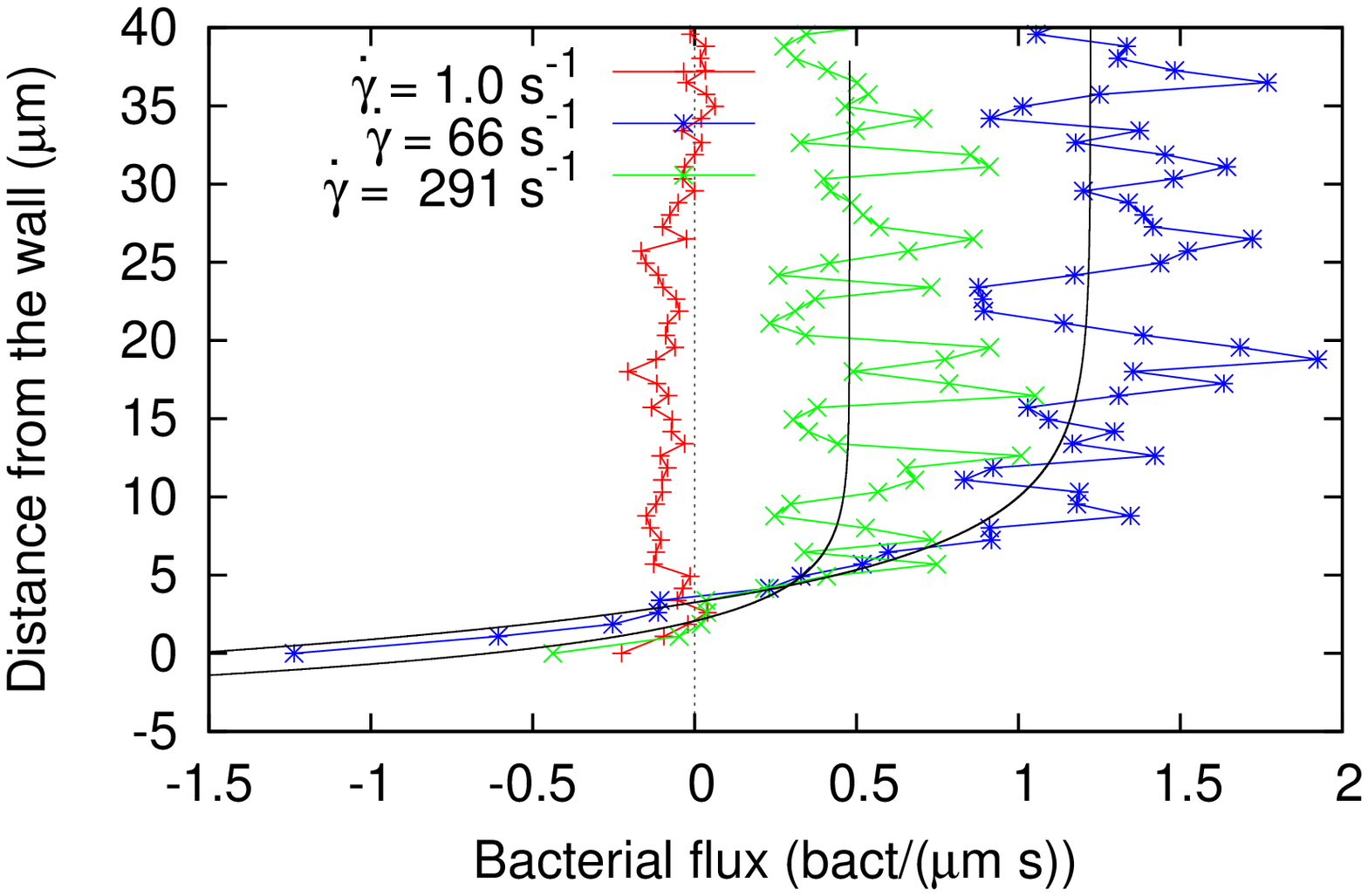} \\
	\flushleft \textbf{b} \\
    \centering  	\includegraphics[width=0.85\linewidth]{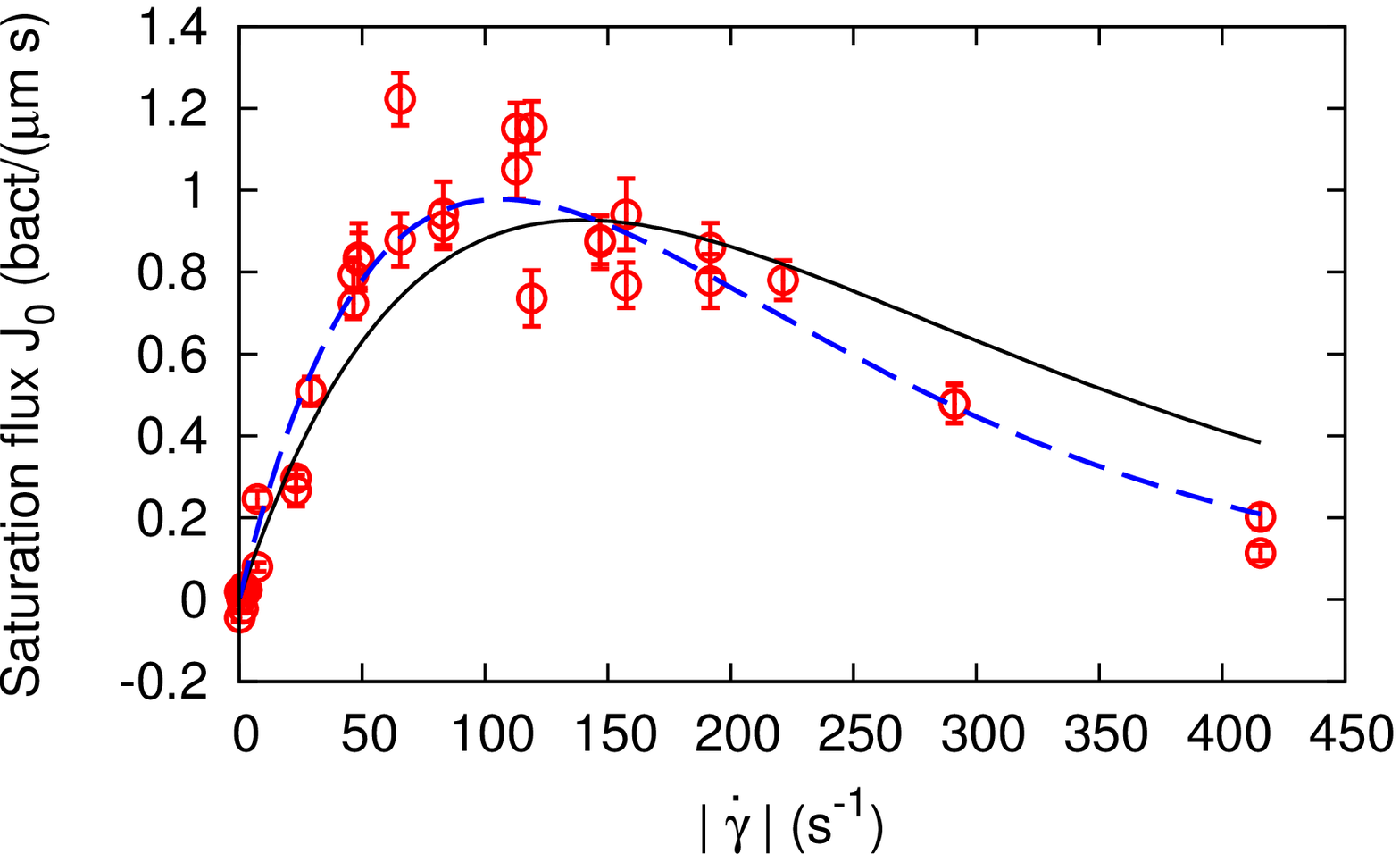} \\
	\flushleft \textbf{c} \\	
  \centering    \includegraphics[width=0.85\linewidth]{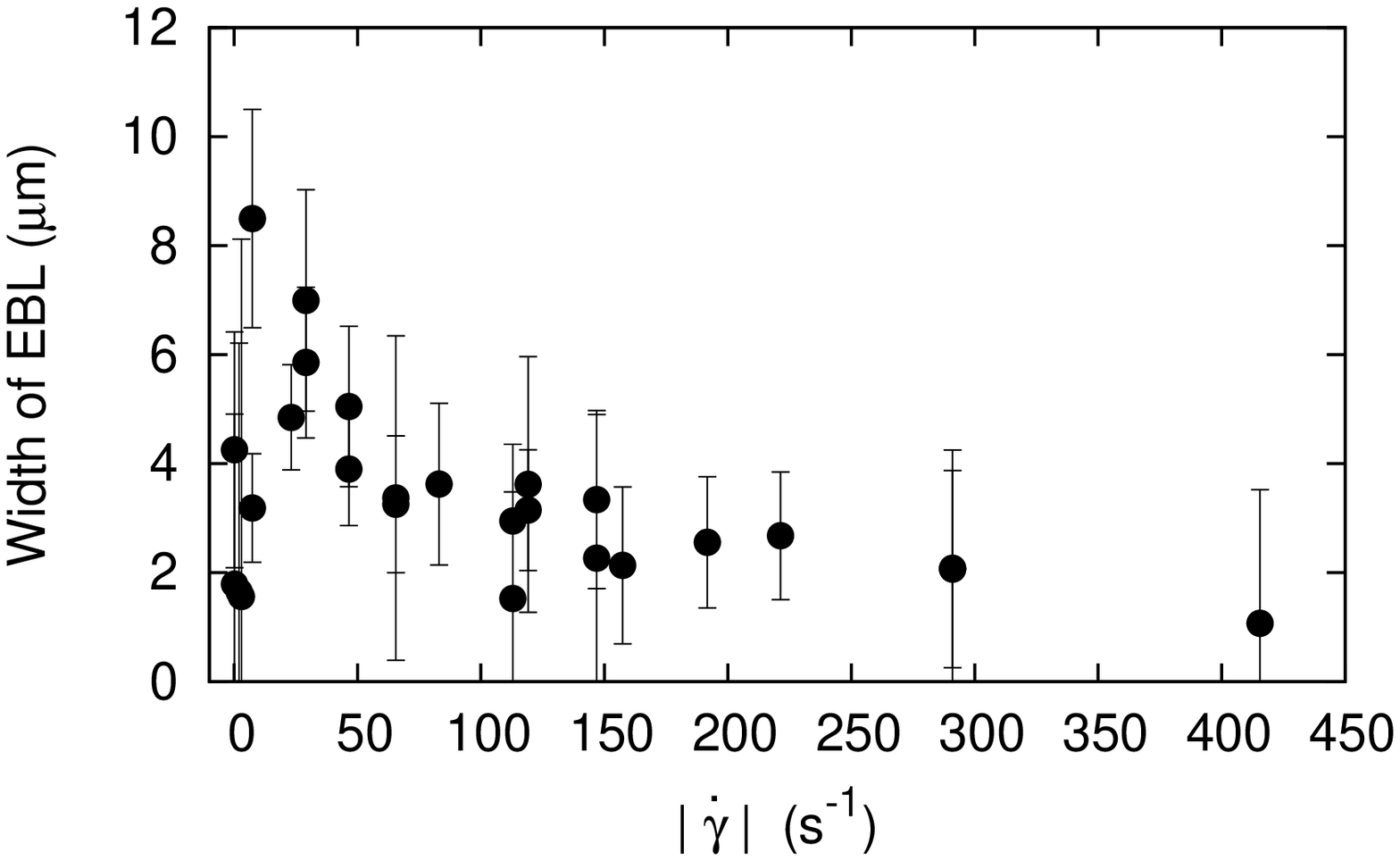}
      \end{center}
  \caption{Edge Boundary layer:
	(a) longitudinal flux density of bacteria $J_x$ as a function of the vertical direction $y$ for 3 values of the shear rate $\dot{\gamma}$.
  For negative values of the horizontal axis bacteria move upstream inside the edge boundary layer, while they move downstream for positive values of the horizontal axis.
	The continuous lines are exponential fits of the form $J_0+B\exp{(-y/\delta)}$ where $J_0$, $B$ and $\delta$ depend on $\dot{\gamma}$.
	(b) Longitudinal Bacteria flux $J_0 $ from the exponential fitting of each bacteria flux vs $\s$. The black solid line is the product $\sigma \left< v_{longitudinal} \right>$ from independent fittings and the blue dashed line is the best  fit to the data using $A \s \exp{-\s/\dot{\gamma}_c}$, where $A = (2.5 \pm 0.3)10^{-2}$bact$/\mu m$ and  $\dot{\gamma}_c = 110 \pm 10 s^{-1}$.
 	(c) Width of the Edge Boundary Layer as a function of the mean shear rate $\dot{\gamma}$.
}
  \label{fig:flux_EBL}
 \end{figure}

The bacteria flux at the bottom surface is represented on Fig. \ref{fig:flux_EBL}(a) as a function of the distance from the wall. If bacteria are transported upstream, the bacteria flux is negative, and it is positive if they are transported downstream. As can be seen in Fig. \ref{fig:flux_EBL}(a), at relatively large distances from the edge, bacteria are advected downstream with the flow and only at very small shear rates (see results for $\s=1s^{-1}$) they move upstream even far away from the edge. The bacteria flux far away from the edges, the saturation flux $J_0$, is represented in Fig. \ref{fig:flux_EBL}(b). When increasing the flow rate, the bacteria flux first increases and then decreases again, illustrating the interplay between bacteria transport and erosion. First the flux increases due to the linear increase in flow velocity (Fig. \ref{fig:vy}(b)), then, the effect of erosion starts to dominate and exponentially decreases the surface concentration (Fig. \ref{fig:n}(a)), leading to an overall decrease of the surface flux. This can be quantified comparing the measured saturation flux $J_0$ with the product $\sigma \left<v_{longitudinal} \right>$ from the expressions obtained by fitting the experimental data in Fig. \ref{fig:n}(a) and Fig. \ref{fig:vy}(b). The result is shown in Fig. \ref{fig:flux_EBL}(b) as the black line. We can also adjust the results for $J_0$ vs the mean shear rate directly with the same functional dependence $J_0=A\s \exp{-\s/\s_c}$. The value obtained for $\s_c$ is within the error bars in agreement with the critical shear rate for erosion at the flat surfaces $\s_s^0$ from Fig. \ref{fig:n}(a). From the fit of the bacteria density and velocity we obtain for $A=1.8 \times 10^{-1}$bact$/\mu m$ slightly smaller than the prefactor obtained from the best fit to the saturation flux.

Close to the edge bacteria are observed to move upstream even at mean shear rates where the flux far from the edges is observed to be downstream. From the longitudinal flux profile, we determine the distance from the edge where the flux changes sign, defining the width of the EBL, which is shown in Fig. \ref{fig:flux_EBL}(c). The boundary layer builds up when the flow is turned on and reaches a maximal width of about  10$\mu m$ at a mean shear rate of about $\s=20s^{-1}$. When further increasing the mean shear rate the EBL decreases again and stabilizes at a value of 2$\mu m$ corresponding to the width of a single bacterium and bacteria move upstream in a single line along the edge.

\section{Conclusions}

In this paper, we have quantified the transport of bacteria at flat surfaces and edges in confined microfluidic channels under flow.

We have measured the decrease of the surface and edge concentrations as a function of applied mean shear rate. The slower decrease of the bacteria concentration at the edge can be explained by the smaller local shear rate at the edge and possibly increased hydrodynamic interactions at the intersection between two flat surfaces. Furthermore bacteria concentrations at different edges are not identical, because rheotaxis at the horizontal surfaces breaks the symmetry along the main axis of the channel, bringing bacteria preferentially towards a given edge: for a given shear flow in one direction, two diagonally opposed edges of the rectangular cross-section are outgoing edges, and the other two are ingoing edges.

We have observed that bacteria swim predominantly upstream at the edges as soon as a small flow is applied. This is attributed to shear induced reorientation of the bacteria attaching to the lateral edges. We have quantified the strong bias in the swimming direction of bacteria at the edges towards upstream
swimming by means of an order parameter that accounts for the
symmetry of up and downstream swimming bacteria concentrations at the edges: the
proportion of bacteria moving upstream increases very quickly as the shear rate
increases until all bacteria are observed to swim upstream.

Bacteria swimming at the edges are nearly undisturbed by the applied flow even at mean shear rates where bacteria transport at the flat surfaces is already strongly influenced by the latter. This can be explained again by the decreased local shear rate at the edges and possibly increased hydrodynamic interactions. Interestingly, we have found that average bacteria velocities along the unidimensional corridor of the edges are mainly the result of collisions between up and downstream swimming bacteria. The number of collisions depends on the total concentration of bacteria at the edges as well as the percentage of up and downstream swimming bacteria and is thus found to be a non monotonic function of the shear rate.

Bacteria are able to swim upstream not only at the edges, but as well within an edge  boundary layer (EBL). The width of the EBL decreases from approximately 10 bacteria
body widths at small shear rates, to 1 body width at higher shear rates. 

Our results thus quantify the bacteria fluxes at all surfaces of a confined microchannel as a function of the mean shear rate. In the future these results can be used to understand bacteria transport in more complex geometries or to design specific flow geometries to guide bacteria fluxes to selected positions.

\bigskip

\textbf{Acknowledgements}
We thank R. Soto and A. Lage-Castellanos for useful discussions. N. F. M. thanks support by the
Pierre-Gilles de Gennes Foundation during PhD studies. E. A. and A. R. acknowledge a TOTAL-ESPCI ParisTech Chair. \\

\bibliographystyle{unsrt}
\bibliography{bibliografia_Alt}

\end{document}